\documentclass[aps,prd,twocolumn,showpacs,groupedaddress,nofootinbib]{revtex4}
\usepackage{graphicx}
\usepackage{color}
\begin{document}

\title{Scalar and Vector 4Q Systems in Anisotropic Lattice QCD}

\author{Mushtaq Loan$^{a}$, Zhi-Huan Luo$^{b}$, and Yu Yiu Lam$^{c}$}
\affiliation{$^{a}$ International School, Jinan University, Huangpu Road West, Guangzhou 510632, P.R. China\\
$^{b}$ Department of Applied Physics, South China Agricultural
University, Wushan Road, Guangzhou, 510642, P.R. China\\
$^{c}$ Department of Physics, Jinan University, Huangpu Road West,
Guangzhou 510632, P.R. China}

\date{November 30,2008}

\begin{abstract}
We present a detailed study of some $4q$ hadrons in quenched
improved anisotropic lattice QCD. Using the $\pi\pi$ and
diquark-antidiquark local and smeared operators, we attempt to
isolate the signal for  $I(J^{P})=0(0^{+}), 2(0^{+})$ and
$1(1^{+})$ states in two flavour QCD. In the chiral limit of
light-quark mass region, the lowest scalar $4q$ state is found to
have a mass, $m^{I=0}_{4q}=927(12)$ MeV, which is slightly lower
than the experimentally observed $f_{0}(980)$. The results from
our variational analysis do not indicate a signature of a
tetraquark resonance in $I=1$ and $I=2$ channels. After the chiral
extrapolation the lowest $1(1^{+})$ state is found to have a mass,
$m^{I=1}_{4q}=1358(28)$ MeV. We analysed the static $4q$ potential
extracted form a tetraquark Wilson loop and illustrated the
behaviour of the $4q$ state as a bound state, unbinding at some
critical diquark separation. From our analysis we conclude that
scalar $4q$ system appears as a two-pion scattering state and that
there is no spatially-localised $4q$ state in the light-quark mass
region.
\end{abstract}
\pacs{11.15.Ha, 12.38.Gc,11.15.Me}

\maketitle

\section{INTRODUCTION}

The concept of multi-quark states has received revived interest
due to the narrow resonances in the spectrum of states. The recent
experimental observation of several new particles as candidates of
multi-quark hadrons are expected to reveal new aspects of hadron
physics \cite{Nakano03,Barmin03,Kubarovsky,Bai04}. Among these
discoveries, the tetraquark systems are also interesting in terms
of their rich phenomenology, in particular the scalar mesons which
still remain a most fascinating subject of research. The five
$0^{++}$ isoscalar mesons below $2$GeV: $f_{0}(400 - 1200),
f_{0}(980), f_{0}(1370), f_{0}(1500)$ and $f_{0}(1760)$ are
subject of conflicting interpretations. $f_{0}(1370)$ is assigned
to be lowest $q\bar{q}$ meson, $f_{0}(1500)$ and $f_{0}(1760)$ are
expected to be the lowest scalar glueball or an $s{\bar s}$ scalar
meson. The differences in the interpretations manifest themselves
in the flavour wave functions of scalar-isoscalar mesons. To
resolve the interpretational differences would require a detailed
analysis of $J/\psi$ decays into vector plus scalar mesons and the
radiative decays like $f_{0}\rightarrow \gamma\omega$ and
$f_{0}\rightarrow \gamma\phi$ that constrains the flavour
structure. Such experiments are feasible at BESII
\cite{BES2a,BES2b,BES2c,ZEUS} and BESIII  with good chances  for
all the channels in radiative $J/\psi$ decays. The tetraquark
systems are also interesting in terms of the nearly stable scalar
charm state $D^{*}_{0}(1980)$, the narrow $D^{*}_{0}(2450)$, the
analogue of $a_{0}(980)$ with open charm, $X(3872)$ and $Y(4260)$
\cite{Belle,CDF,Barbar,Close03,Close04}. From the consideration of
their masses, narrow decay width  and decay pattern, these states
cannot be regarded as the simple mesons of $c{\bar c}$ or $c{\bar
s}$ but conjectured to be $4q$ or molecular states
\cite{Pakvasa,Wong,Braaten}.

While a phenomenological model may often be helpful in obtaining a
qualitative understanding of the data, lattice QCD, at the present
status of approximations, seems to provide a trustworthy guide
into unknown territory in multi-quark hadron physics
\cite{Lasscock,Oikharu}. It is expected to play an important role
in revealing the real nature of multi-quark hadrons. There are
several, mainly exploratory, lattice studies of tetraquark systems
available
\cite{Fukugita,Mark,Gupta,Sharpe,Hideo,Alexandrou,Fumiko,Hideo07},
with mixed results. Most of these calculations claim the existence
of a bound scalar $I=0$ tetraquark state whereas Ref.
\cite{Hideo07} has observed no evidence for a $4q$ resonance. One
may naively interpret the negative results obtained in Ref.
\cite{Hideo07} as a consequence of their combined analysis of
maximum entropy method with hybrid boundary condition method on
their choice of operators, which would strongly couple to two-pion
scattering states.  Thus, it is important to use various
interpolating operators and variational techniques on more
accumulated data to make a precise prediction. Our approach will
be to use a number of different interpolating fields and smeared
operators to enhance the low-lying spectra.

In this study, we provide a detailed analysis on tetraquark
hadrons from the combination analysis of variational method and
smearing on various interpolating fields on improved anisotropic
lattices. Using the quenched approximation, and discarding
quark-antiquark annihilation diagrams, we construct $q^{2}{\bar
q}^{2}$ sources from multiple operators. We exclude the processes
that mix $q{\bar q}$ and $q^{2}{\bar q}^{2}$ and allow the quark
masses to vary from small to large values. In the absence of quark
annihilation, we do not expect any mixing of $q^{2}\bar{q}^{2}$
with pure glue. Thus we can express the $q^{2}\bar{q}^{2}$
correlation functions in terms of a basis determined by gluon and
quark exchange diagrams only. In the process of searching for a
bound state it is essential to explore a large number of
interpolating fields having the quantum numbers of the desired
state. Explicitly, one needs to construct an interpolating field
which has significant overlap with the $4q$ system. However, any
$(qq{\bar q}\bar{q})$ operator must couple to hadronic states with
the same quantum numbers (for example, the $\pi\pi$ scattering
state). It is necessary to disentangle the lowest-lying $4q$
states from the $\pi\pi$ scattering states, as well as the excited
$4q$ states. To this end, we adopt a variational method to compute
a $n\times n$ correlation matrix from the interpolating fields and
from its eigenvalues we extract the masses.

The rest of the paper is organized as follows:  in Sec. II, we
discuss our choice of interpolating fields for the supposed scalar
tetraquark state and outline our  construction of the correlation
matrix analysis.  The technical details of the lattice simulations
as well as the actions used in this study are discussed in Sec.
III. It is essential to use the improved action that displays
nearly perfect scaling, since large scaling violations could lead
to a false signature of attraction. We therefore perform our
simulations on improved lattice actions with tadpole-improved
renormalised parameters that provide continuum limit results at
finite-lattice spacing. We present and discuss our numerical
results in Sec. IV. Finally, we  perform the study of the
inter-quark potential in our $4q$ system to establish a connection
between connected $4q$ state and the ``two-meson'' state around
the cross-over. Sec. V is devoted to our summary and concluding
remarks.

\section{Lowest $q^{2}{\bar q}^{2}$ states on the Lattice}

We propose interpolators designed to maximize the possibility to
observe attraction between tetraquark constituents at relatively
heavy quark masses. Two general types of operators are considered:
those based on the $\pi \pi$ configuration, and those based on a
diquark-antidiquark configuration. The simplest $\pi\pi$-type
interpolators  for the $I=0$ and $I=2$ channels, respectively, are
\begin{eqnarray}
O_{1}^{I=0}(x)&  =&  \left[\big\{\big({\bar d}^{a}(x)\gamma_{5}
u^{a}(x)\big)\big({\bar u}^{b}(x)\gamma_{5} d^{b}(x)\big)\right.
\nonumber\\
& & \left. -(u\leftrightarrow d, {\bar u}\leftrightarrow {\bar
d})\big\} + \frac{1}{2}\big\{\big({\bar u}^{a}(x)\gamma_{5}
u^{a}(x)\big) \right.
\nonumber\\
& & \left. \times \big({\bar d}^{b}(x)\gamma_{5} d^{b}(x)\big)-
(u\leftrightarrow d, {\bar u}\leftrightarrow {\bar
d})\big\}\right],
\end{eqnarray}
and
\begin{equation}
O_{2}^{I=2}(x)  =  \big({\bar d}^{a}(x)\gamma_{5}
u^{a}(x)\big)\big({\bar d}^{b}(x)\gamma_{5} u^{b}(x)\big).
\end{equation}

The $J^{P}=1^{+}$ operator with isospin $I=1$ is evaluated from
the pseudo-scalar and vector mesons fields and has the form:
\begin{eqnarray}
O_{3}^{I=1}(x) & =&  \frac{1}{2}\left[({\bar
d}^{a}(x)\gamma_{i}u^{a}(x))({\bar d}^{b}(x)\gamma_{5}u^{b}(x))
\right.
\nonumber\\
& & \left. - ({\bar d}^{a}(x)\gamma_{5}u^{a}(x))({\bar
d}^{b}(x)\gamma_{i}u^{b}(x)) \right.
\nonumber\\
& & \left. +({\bar d}^{a}(x)\gamma_{i}u^{a}(x))({\bar
d}^{b}(x)\gamma_{i}u^{b}(x))\right].
\end{eqnarray}
The other type of interpolating field is one in which quarks and
antiquarks are coupled into a set of diquark and antidiquark,
respectively and has the form
\begin{equation}
O_{4}(x)  = \epsilon_{abc}[u^{T}_{a}C\Gamma
d_{b}]\epsilon_{dec}[{\bar u}_{d}C\Gamma {\bar d}^{T}_{e}].
\end{equation}
Accounting for both colour and flavour antisymmetry, the possible
$\Gamma$s are restricted to within $\gamma_{5}$ and $\gamma_{i}$.
The operators $O_{1,2}$ are similar to the naive \mbox{pion
$\otimes$ pion} operator with the difference that in the former
one has a contribution from gluon and quark exchange diagrams to
the $\pi - \pi$ four-point functions. The operator $O_{4}$ is
motivated by the Jaffe-Wilzcek description of diquarks
\cite{Jaffe03}, which is expected to have a small overlap with
two-meson scattering states.

\subsection{Extraction of masses}
The mass of the ground state is extracted from the asymptotic
behaviour of the two-point the temporal $4Q$ correlator
\begin{equation}
C(t) = \frac{1}{V}\sum_{\vec x}\langle O ({\vec x}
,t)O^{\dagger}({\vec 0},0)\rangle ,
\end{equation}
where the total momentum of the $4q$ system is projected to be
zero. To disentangle the lowest-lying $4q$ states from $\pi\pi$
scattering states this way, we adopt a variational method to
compute a $2\times 2$ correlation matrix from two different
interpolating fields.

We compute the propagators $\langle (O_{i})_{x}({\bar
O}_{j})_{y}\rangle$ with fixed $y=({\vec 0},0)$, and their time
correlation functions,
\begin{equation}
C_{ij}(t) =\langle \sum_{ \vec x}\mbox{tr}\left[\langle
(O_{i})({\vec x},t){\bar O}_{j})({\vec
0},0)\rangle_{f}\right]\rangle_{U},
\end{equation}
where the trace sums over the Dirac space, and the subscripts $f$
and $U$ denote fermionic average and gauge field ensemble average,
respectively. Following \cite{Lasscock,Alexandrou05,Ting05} we
solve the eigenvalue equation
\begin{equation}
C(t_{0})v_{k}(t_{0})=\lambda_{k}(t_{0})v_{k}(t_{0})
\end{equation}
to determine the eigenvectors $v_{k}(t_{0})$ and use these
eigenvectors to project the correlation matrices to the space
corresponding to the $n$ largest eigenvalues $\lambda_{n}(t_{0})$
\begin{equation}
C_{ij}^{n}(t) = (v_{i},C(t)v_{j}), \hspace{1.0cm} i,j=1, \cdots ,n
\end{equation}
and solve the generalised eigenvalue  equation for the projected
correlation matrix $C_{ij}^{n}$.  In practice, we extract our
results from a $2\times 2$ correlation matrix. The masses are
extracted from the average of $C_{ij}$ by a single-exponential fit
to $\lambda_{i}(t)$ for the range of $t$ in which the effective
mass
\begin{equation}
M_{eff} = \ln\left[\frac{\lambda (t)}{\lambda (t+1)}\right]
\label{eqn9}
\end{equation}
attains a plateau. However, to ensure the validity of our results,
we compare them with those obtained using
\begin{equation}
M^{'}_{eff} = \ln\left[\frac{\lambda (t-1)-\lambda (t)}{\lambda
(t)- \lambda (t+1)}\right]\label{eqn10}.
\end{equation}

\subsection{Static $4q$ potential}

The static tetraquark potential is extracted from the gauge
invariant $4q$ Wilson loop. The SU(3) Wilson loop is constructed
by creating a $4q$ state at a time $t=0$, which is annihilated at
a later time $t$. Following \cite{Alexandrou,Fumiko} we write the
$4q$ Wilson loop as
\begin{eqnarray}
W_{4q} & = &
\frac{1}{12}\epsilon^{abc}\epsilon^{def}\epsilon^{a'b'c'}\epsilon^{d'e'f'}
U({\bf x,x}',1)^{aa'}U({\bf x,x}',2)^{bb'} \nonumber\\
& & \times U_{G}({\bf x,y})^{cf}U({\bf y}',{\bf y},3)^{d'd}U({\bf
y}',{\bf y},4)^{e'e} \nonumber\\
& &  \times U_{G'}({\bf y}',{\bf x}')^{f'c'}
\end{eqnarray}
where the staple-like links $U({\bf x,x}',k)$ are given by
\begin{displaymath}
U({\bf x,x}',k) = P\exp
\left[ig\int_{\Gamma_{k}}dz^{\mu}A_{\mu}(z)\right],
\end{displaymath}
where $\Gamma_{k}$ denotes the path from ${\bf x}$ to ${\bf x}'$
for quark line $k$.

\begin{figure}[!h]
\scalebox{0.90}{\includegraphics{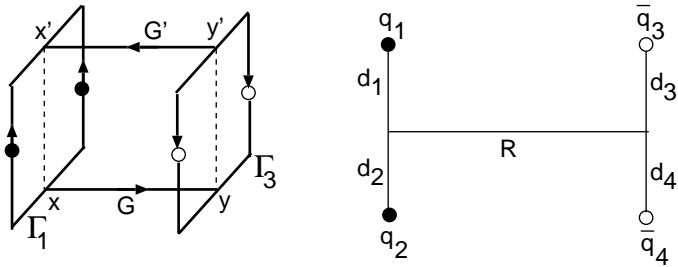}} \caption{
\label{fig0} The Wilson loop for the tetraquark system.}
\end{figure}

The junctions at ${\bf x}$ and ${\bf y}$ are joined by a line-like
contour
\begin{displaymath}
U_{G}({\bf x,y}) = P\exp \left[ig\int_{G}d{\bf z} \cdot {\bf
A}(z)\right],
\end{displaymath}
at $t=0$. The tetraquark potential can be obtained from the large
$t$ behaviour of $\langle W_{4q}\rangle$,  where the ground-state
contribution becomes dominant. To obtain the optimal
signal-to-noise ratio, we must suppress the contributions from the
excited states, which may be done by  APE smearing technique
\cite{APE87}. This involves replacing the space-like link
variables $U_{i}$ by smeared link variables. To tune the smearing
parameters, we fix the smearing fraction to $\alpha =0.7$, as it
is sufficient to fix $\alpha$ and tune  the smearing level to
$n_{APE}$. A typical value which enhances the ground-state
component in the $4q$ Wilson loop was $n_{APE}=8$. The tetraquark
potential is then extracted in the standard way from the large $t$
behaviour of the $4q$ Wilson loop with a single-exponential fit:
\begin{equation}
\langle W_{4q}\rangle = Ze^{-V_{4q}t}
\end{equation}
in the range $[t_{min},t_{max}]$. The effective potential is then
extracted from
\begin{equation}
V_{4q}(t) \equiv \ln\left[ \frac{W_{4q}(t)}{W_{4q}(t+1)}\right].
\end{equation}
In this way, we calculate the potential for our planar $4q$
configurations.

\section{Simulation details}

To examine  $q^{2}\bar{q}^{2}$ in lattice QCD, we explore the
improved actions on anisotropic lattices. With most of the
finite-lattice spacing artifacts having been removed, one can use
coarse lattices, with fewer sites and much less computational
effort. It has been found that even on fairly coarse lattices
these actions give good results for hadron masses by estimating
the coefficients of the improvement terms using
tadpole-improvement. Using a tadpole improved anisotropic gluon
action \cite{Morningstar99}, we generate quenched configurations
on $16^{3}\times 64$ lattice (with periodic boundary conditions in
all directions) at $\beta = 4.0$. Gauge configurations are
generated by a 5-hit pseudo heat bath update supplemented by four
over-relaxation steps. These configurations are then fixed to the
Coulomb gauge at every 500 sweeps. After discarding the initial
sweeps, a total of 300 configurations are accumulated for
measurements. Quark propagators are computed by using a
tadpole-improved clover quark action on the anisotropic lattice
\cite{Okamoto02}. All the coefficients in the action are evaluated
from a tree-level tadpole improvement. We calculate the spectrum
at quark masses as light as $0.004$, which is much less than the
hadronic scale, $1.0$ GeV, associated with the scalar $4q$ system.

On each gauge fixed configuration, we invert the quark matrix by
the BiCGStab algorithm to obtain the quark propagator. Using
anti-periodic boundary conditions in the time direction, we
calculate the propagators on three source time slices on the same
lattice. For the lattice analysed here, we place the wall sources
on the first, second and third time slices.  Since there is a
possibility of correlation among the propagators with different
source time slices, we average them and treat them as a single
result in a jackknife analysis.

The long distance exponential falloff is determined by the
lowest-energy hadron state to which the operator couples. For
lattice QCD, one has options to use a local or nonlocal
interpolating operator for hadrons, provided its correlation
function has a significant overlap with the state under
consideration. To obtain a better overlap with the ground state,
we used iterative smearing of gauge links and the application of
the fuzzing technique for the fermion fields \cite{UKQCD1995}
taking the physical size of the particle into account. The fuzzed
quark field is constructed to be symmetric in all space
directions. Such a fuzzed quark field is only used at the sink,
while the one at the source remains local. The application of
fuzzing for two of the four quarks inside the $q^{2}{\bar q}^{2} $
flattens the curvature of the effective mass. The largest plateau
in the region with small errors is obtained with fuzzed $u$- and
$d$-quarks. We used this variant to calculate our correlation
functions.

We estimate the lattice spacing by linearly extrapolating the
$\rho$ mass to the physical quark mass. The latter can be
determined from the ratio of two non-strange hadrons. Since
$m_{u}$ and $m_{d}$ are fixed by the nucleon mass, in the
non-strange sector, $m_{\rho}$ is a test both for finite-size
effects and quenching errors. The major source of discrepancy
among the lattice spacings from different observables is the
quenching effect. The obtained $\rho$ and $N$ masses are compared
to the experimental values and show a deviation of less than
$3-4\%$ for the lattice size explored here. Such a variance can be
considered as the usual quenching effect. Inspired by the good
agreement of the $\rho$ and $N$ masses with the experimental
values, the scale was set alternatively by the ratio
$m_{\pi}/m_{N}$. Using the experimental value $768$ MeV for the
rho mass, the spacing of our lattice is $a_{s}=0.462(2)$ fm. The
bare quark mass is determined by extracting the mass of the $K$
meson. At $m_{q}a_{t}=0.04$, we obtained the mass of vector meson
$m_{K}= 553(2)$ MeV, which is in good agreement with the estimates
obtained from other studies \cite{Ishii,Alexandrou05}. Thus taking
the strange quark bare mass to be $m_{s}a_{t}=0.04$, we choose the
bare quark masses for $u$ and $d$ to be much smaller than $m_{s}$,
i.e., $m_{q}a_{t}=0.01, 0.008, 0.0065, 0.0055, 0.005, 0.004$.

\section{Results and discussion}

In this section we present the results of our lattice simulation
of tetraquark masses for the isospin $I=0, 1$ and $I=2$ channels.
In addition to extracting the masses of the $q^{2}\bar{q}^{2}$
states, we also study the mass differences between the candidate
tetraquark states and the free two-particle states. This analysis
is actually important in determining the nature of the states
observed on the lattice, and the identification of the true
resonances. Hence a lattice signature which might be observed for
tetraquark resonance is a negative mass splitting at quark masses
near the physical regime.

\begin{figure}[!h]
\scalebox{0.45}{\includegraphics{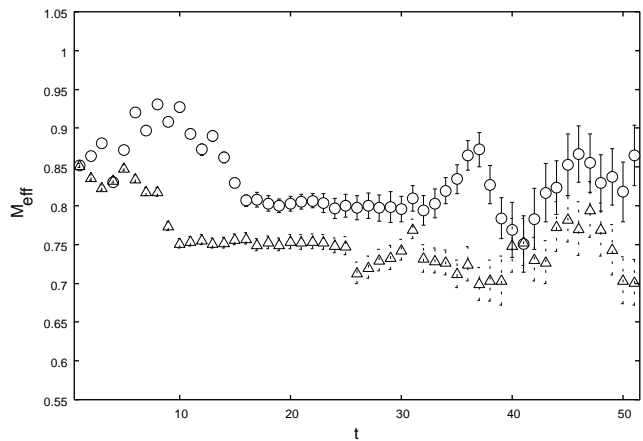}} \caption{
\label{fig1} Effective mass of $4q$ in the $I=0$ (open triangles)
and $I=2$ (open circles) channels at $\kappa_{t}=0.2635$
($m_{q}a_{t}= 0.0065$).}
\end{figure}

Fig. \ref{fig1} shows the effective mass plot for the lowest
scalar $4q$ $(ud\bar u\bar d)$ for the typical hopping parameter
$\kappa_{t}=0.2635$. At sufficiently large $t$, contributions from
excited states are diminished, and the correlator is dominated by
a single state. Thus a plateau may serve as an indicator of the
single-state saturation, hence allowing one to perform the
single-exponential fit to the plateau region. In the region $0\leq
t \leq 9$, the effective mass for the $I=0$ channel decreases
monotonically and  reaches a stable plateau  at $10\leq t\leq 25$.
Beyond $t \sim 25$ the data become noisy. The fitting range
$[t_{min},t_{max}]$ for the final analysis is determined by fixing
$t_{max}$ and finding the range of $t_{min}$ where the
ground-state mass is stable against $t_{min}$.
We choose one ``best fit" which  is insensitive to the fit range,
has high confidence level and reasonable statistical errors. We
then confirm this by looking at the plateau region of the
correlator. Statistical errors of masses are estimated by the
jackknife method. Since the analysed configurations were highly
uncorrelated, which we ensured by separating the analysed
configurations by as many as 1000 sweeps, the statistical errors
of our mass estimates are typically on the few percent level. With
much of our data we find that the errors are purely statistical,
and the goodness of the fit is gauged by the $\chi^{2}/N_{\rm
DF}$. All fits are reasonable, having $\chi^{2}/N_{\rm DF} \leq
1$. For the $I=2$ channel it was possible to find a fit region
$[t_{15},t_{35}]$ in which a convincing plateau was observed. The
effective mass is found to be stable using different values of $t$
in Eq. (\ref{eqn9}), which suggests that the tetraquark ground
state is correctly projected.

\begin{figure}[!h]
\scalebox{0.45}{\includegraphics{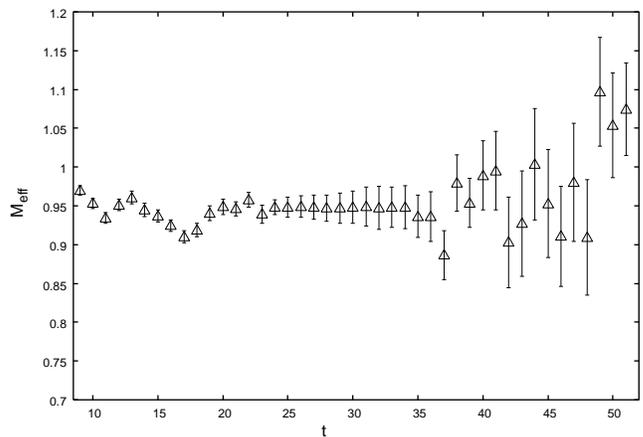}} \caption{
\label{fig2} Effective mass of $I(J^{P})=1(1^{+})$ $4q$ state  at
$\kappa_{t}= 0.2650$ $(m_{q}a_{t}=0.005$).}
\end{figure}

For the $I(J^{P})=1(1^{+})$ tetraquark state, Fig. \ref{fig2}, we
observe that the exponential seems to oscillate in time at the
lowest-quark mass at large $t$. One can still demonstrate the
existence of a plateau in the effective mass and extract the mass
with some reliability, although the errors are relatively large.
Suppressing any data point which has an error larger than its mean
value, a possible plateau is seen in the region $20\leq t\leq 35$
with reasonable errors, where the single-state dominance is
expected to be achieved. To ensure the validity of our results, we
compared them to those obtained using Eq. (\ref{eqn10}). It was
found that the evaluations of (\ref{eqn9}) and (\ref{eqn10}) yield
results very consistent within statistical errors.

\begin{figure}[!h]
\scalebox{0.45}{\includegraphics{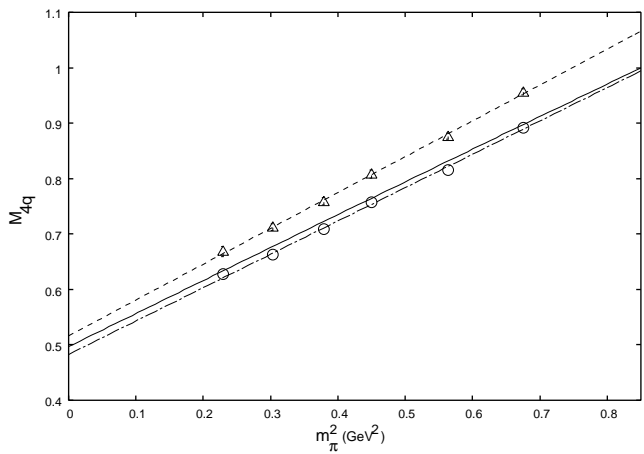}} \caption{
\label{fig3} Chiral extrapolation of effective masses in the $I=0$
(open circles) and $I=2$ (open triangles) channels. The solid
curve denotes the two-pion threshold, and dashed curves are the
linear fits to the data.}
\end{figure}

The chiral extrapolation to the physical limit is the next
important issue. From the view point of chiral perturbation
theory, data points with smallest $m_{\pi}^{2}$ should be used to
capture the chiral log behaviour. Leinweber et al. \cite{Derek04}
demonstrated that the chiral extrapolation method based upon
finite-range regulator leads to extremely accurate values for the
mass of the physical nucleon with systematic errors of less than
one percent. We use a set of data points with smallest
$m_{\pi}^{2}$ to capture the chiral log behaviour.  Fig.
\ref{fig3} collects and displays the resulting particle masses
extrapolated to the physical quark mass value using linear and
quadratic fits in $m_{\pi}^{2}$,
\begin{displaymath}
m_{h}=a+bm_{\pi}^{2},\hspace{0.10cm}
m_{h}=a+bm_{\pi}^{2}+cm_{\pi}^{4}.
\end{displaymath}
The difference between these two extrapolations gives some
information on the systematic uncertainties in the extrapolated
quantities. Although, our quark masses are quite small, both
linear and quadratic fits essentially gave the identical results.
The chiral uncertainties in the physical limit are significantly
small (less than a percent) at our present statistics.

At smaller quark masses, the data for the $I=0$ channel appear
slightly below the two-pion threshold, by $\sim 30$ MeV, and seems
to behave linearly in $m_{\pi}^{2}$ rather than the two-pion
threshold which shows a quadratic behaviour in $m_{\pi}^{2}$. The
data for $I=2$ lie  above the two-pion threshold by $\sim 40-50$
MeV for all quark masses analysed here.  If the slight difference
from two-pion threshold can be explained by the the two-pion
interaction, then the lowest scalar $4q$ state can be regarded as
a two-pion scattering state.

\begin{figure}[!h]
\scalebox{0.45}{\includegraphics{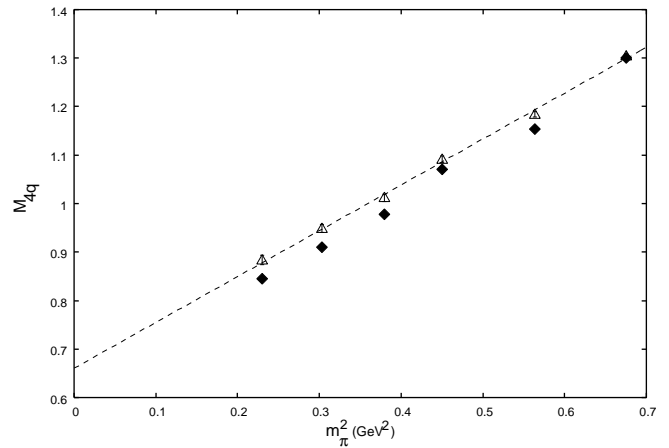}} \caption{
\label{fig4} Chiral extrapolation of effective masses in
$I(J^{P})=1(1^{+})$. The open triangles denote the $4q$ data and
the solid diamonds  denote the two-particle state ($\pi +\rho$).
The dashed curve is a linear fit to the data.}
\end{figure}

The results for the extracted mass for $J^{P}=1^{+}$ are displayed
in Fig. \ref{fig4} as a function of $m^{2}_{\pi}$. The
ground-state masses for the tetraquark and two-particle states are
again very different. The mass of the tetraquark state extracted
is consistently higher than the lowest two-particle state, namely
 $\pi +\rho$. This trend continues in the physical limit where
the masses exhibit the opposite behaviour to that which would be
expected in the presence of binding. In the chiral limit the $I=1$
state lies above the two-pion threshold by $\sim 70$ MeV.

\begin{table}[!h]
\caption{ \label{tab1} The masses of the $I(J^{P})=0(0^{+}),
2(0^{+})$ and $I(J^{P})=1(1^{+})$ $4q$ states in the lattice units
for various values of $\kappa_{t}$.}
\begin{ruledtabular}
\begin{tabular}{ccccccc}
   &   & \multicolumn{2}{c}{$J^{P}=0^{+}$} & $J^{P}=1^{+}$\\
$\kappa_{t}$ & $M_{\rho}$ & $M^{I=0}_{4q}$ & $M^{I=2}_{4q}$ &
$M^{I=1}_{4q}$\\ \hline
0.2610  & 0.850(2) & 0.892(4) & 0.956(6)  & 1.305(4)  \\
0.2620  & 0.741(2) & 0.815(5) & 0.877(6)  & 1.174(6)  \\
0.2635  & 0.687(3) & 0.758(5) & 0.808(7)  & 1.094(7)  \\
0.2640  & 0.606(3) & 0.709(6) & 0.756(7)  & 1.012(7) \\
0.2650  & 0.572(3) & 0.663(7) & 0.712(6)  & 0.949(8)  \\
0.2655  & 0.529(4) & 0.627(8) & 0.668(8)  & 0.884(9) \\
\end{tabular}
\end{ruledtabular}
\end{table}

Since quenched spectroscopy is quite reliable for the mass ratio
of stable particles, it is physically even more motivated to
extrapolate the mass ratio instead of the mass. This allows for
the cancellation of systematic errors since the hadron states are
generated from the same gauge field configurations and hence
systematic errors are strongly correlated. The mass ratios at hand
show a remarkably small scaling violation; -hence, we adopt an
$m_{\pi}^{2}$-linear extrapolation for the continuum limit.  We
also perform an $m_{\pi}$-quadratic extrapolation to estimate
systematic errors. Performing such extrapolations for all sets of
masses, we adopt the choice which shows the smoothest scaling
bahaviour for the final value, and we use others to estimate the
systematic errors.
\begin{figure}[!h]
\scalebox{0.45}{\includegraphics{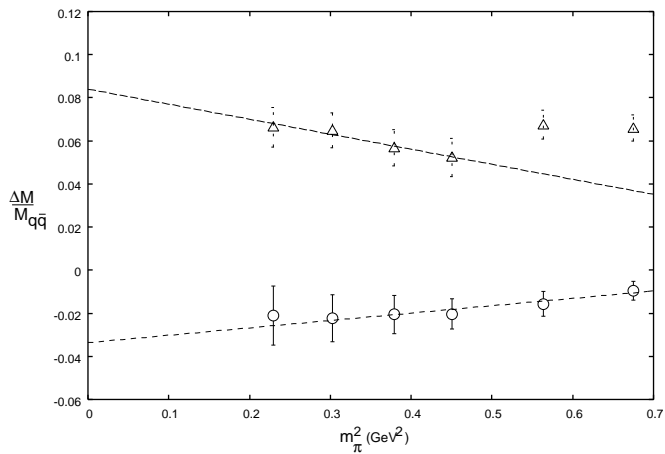}}
\caption{\label{fig5}Extrapolation of the mass ratios of the $I=0$
(open circles) and $I=2$ (open triangles) channels to the physical
limit. The dashed lines are the linear fits in $m^{2}_{\pi}$ to
the data.}
\end{figure}

The  results of the mass difference analysis are shown in Table
\ref{tab2} and illustrated in Fig. \ref{fig5}. The data behave
almost linearly in $m_{\pi}^{2}$ and both  linear and quadratic
fits essentially gave  identical results. Again the contributions
from the uncertainties due to chiral logarithms in the physical
limit are seen to be significantly less dominant.

The question whether the lowest-lying scalar $4q$ state extracted
is a scattering state or a resonance is better resolved by
analysing the ratio of the mass differences $\Delta M =
M_{4q}-2M_{q{\bar q}}$ (between the candidate lowest scalar $4Q$
state and the two-pion state) and the $\rho$ mass.

The non-zero $m^{2}_{\pi}$ values of the ratio are within
0.01-0.02 standard deviations of the extrapolated zero pion mass
result. This implies that the chiral uncertainties in the physical
limit are less than $2\%$. The fitted results show that the mass
difference $\Delta M$ and hence the mass ratio for the $I=0$
channel are clearly negative and increase in magnitude as we
approach the physical regime. This implies that the binding
becomes stronger at light-quark masses with a general trend of
negative binding as the zero quark mass limit is approached. In
the physical limit, the negative mass difference is $\sim
25-30(2)$ MeV. Naively one may be tempted to interpret this
negative mass splitting near the physical regime as a lattice
resonance signature for scalar $4q$ resonance with $I=0$. Using
the chirally extrapolated value of the ratio
$m_{\pi}/m_{\rho}=0.60(1)$, the $I=0$ state is found to have a
mass of $m^{I=0}_{4q}=927(12)$ MeV in the physical limit. The
lowest scalar $4q$ state appears to be slightly lighter than the
experimentally observed $f_{0}(980)$. It still remains to verify
whether analysis at relatively large quark masses would affect the
manifestation of the bound state and aid to confirm the indication
of the resonance. This analysis is discussed in the next section.
On the other hand the $I=2$ channel again shows a positive
splitting of the order of $\sim 50(4)$ MeV. This suggests that
instead of a bound state, we appear to be seeing a scattering
state in the $I=2$ channel.

\begin{figure}[!h]
\scalebox{0.45}{\includegraphics{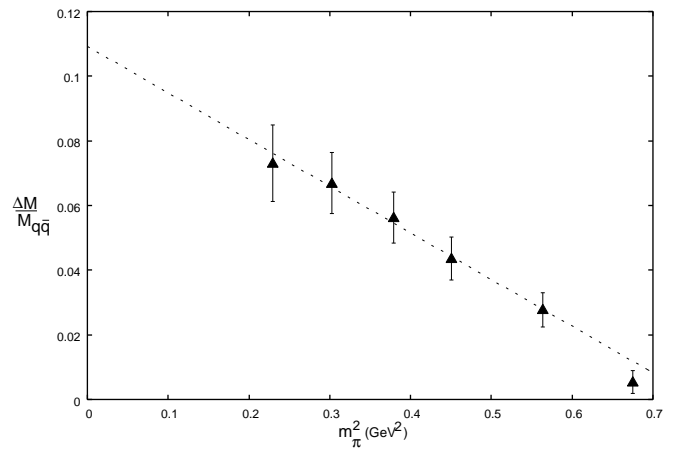}} \caption{
\label{fig6} As in Fig. \ref{fig5}, but for the
$I(J^{P})=1(1^{+})$ state.}
\end{figure}

The behaviour observed for the mass differences between the $I=1$
and the two-particle states is illustrated in Fig. \ref{fig6}. A
positive mass splitting is observed for all the five smallest
quark masses and increases  as the physical regime is approached.
The signature of repulsion at quark masses near the physical
regime would imply no evidence of the resonance in the $I=1$
channel. This suggests that the $I=1$ tetraquark system (in the
quenched approximation) is more complicated and different from the
two-hadron system otherwise. Adopting a linear chiral
extrapolation, the mass of the $I(J^{P})=1(1^{+})$ tetraquark
state is estimated as $M^{J=1}_{4q}= 1360(30)$ MeV which is lower
than the mass of the experimentally observed $a_{0}(1474)$.

\begin{table}[!h]
\caption{ \label{tab2} Ratios between the mass difference, $\Delta
M_{4q}$ and  $m_{\rho}$ at various pion masses. }
\begin{ruledtabular}
\begin{tabular}{ccccccc}
$M_{\pi}(GeV)$ & $\Delta M^{I=0}_{4q}/M_{\rho}$ &
$\Delta M^{I=2}_{4q}/M_{\rho}$ & $\Delta M^{I=1}_{4q}/M_{\rho}$ \\
\hline
0.822  & -0.0096(43) & 0.0660(61) & 0.0054(36) \\
0.751  & -0.0016(54) & 0.0674(67) & 0.0278(53)\\
0.672  & -0.0203(71) & 0.0523(86) & 0.0436(66)   \\
0.616  & -0.0206(88) & 0.0568 (82)& 0.0562(78) \\
0.550  & -0.022(10)  & 0.0650(81) & 0.0669(91)   \\
0.479  & -0.021(13)  & 0.0664(92) & 0.0731(94) \\
\end{tabular}
\end{ruledtabular}
\end{table}

\begin{figure}[!h]
\scalebox{0.45}{\includegraphics{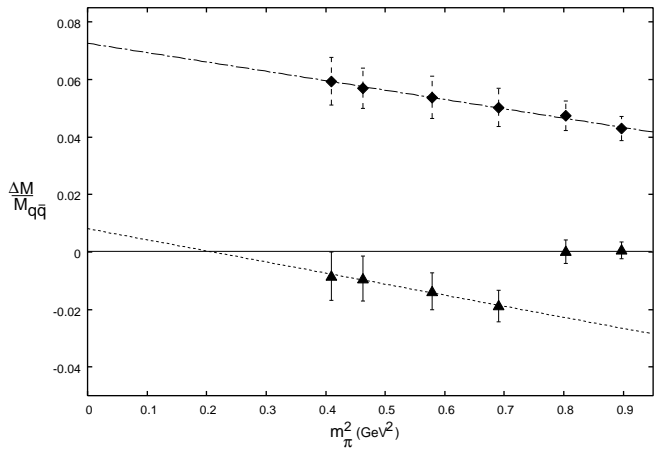}} \caption{
\label{fig7} Chiral extrapolation of the mass ratios at heavy
quark masses. The solid triangles denote the $I=0$, and solid
diamonds represent the $I=1$ tetraquark states, respectively. The
dashed curves are linear fits to the data.}
\end{figure}

As mentioned above, to obtain a definitive result for the
signature of a possible scalar tetraquark bound state with $I=0$
will require the implementation of the heavy quark limit. The
heavy quark mass suppresses relativistic effects, which
complicates the interpretation of light-quark states. By giving
the quarks a larger mass would alter threshold, which in turn
would affect the manifestation of the bound state. To confirm
this, we calculate the mass of the lowest $4q$ state with varying
quark mass. We allow the quark mass to be larger (hundreds of MeV)
so that the continuum threshold for the decay $q^{2}{\bar
q}^{2}\rightarrow (q{\bar q})(q{\bar q})$ is elevated. The
resulting extracted masses and mass differences are tabulated in
Table \ref{tab3} and shown in Fig. \ref{fig7}, respectively.

The ground-state masses in the scalar channel are slightly larger
than those obtained at the small quark masses. The mass difference
is consistent with or slightly below the mass of the lowest
two-pion state at the four smallest quark masses. The mass
difference appears to be approaching a positive constant near the
physical limit and the tetraquark masses are $\sim 50$ MeV larger
than the $s$-wave two-pion state. This suggests that instead of a
bound state, we appear to be seeing a scattering state in the
$I=0$ channel. The lattice data for the $I=1$ channel appear to be
well above the increased two-particle state threshold and remains
approximately constant with $m_{\pi}^{2}$.  In the chiral limit,
the  difference between the $I=1$ $4q$ state and the two-pion
threshold is $\sim 80$ MeV. Again, the positive mass difference
could be a signature of repulsion in this channel. We conclude
that there are no surprises in the $I=1$ channel -no evidence of
attraction and hence no indication of a resonance which could be
interpreted as the $4q$ state.

\begin{table}[!h]
\caption{ \label{tab3} Ratios between the mass difference, $\Delta
M_{4q}$ and the $m_{\rho}$ at large quark masses. }
\begin{ruledtabular}
\begin{tabular}{ccccccc}
$M_{\pi}(GeV)$ & $\Delta M^{J=0}_{4q}/M_{\rho}$ & $\Delta
M^{I=1}_{4q}/M_{\rho}$  \\ \hline
0.947  &   0.0006(30) &  0.0429(42) \\
0.896  &   0.0002(41) &  0.0474(51) \\
0.830  &  -0.0188(55) &  0.0503(67) \\
0.761  &  -0.0137(64) &  0.0538(74) \\
0.680  &  -0.0093(78) &  0.0570(70) \\
0.641  &  -0.0085(84) &  0.0594(82)\\
\end{tabular}
\end{ruledtabular}
\end{table}

\begin{figure}[!h]
\scalebox{0.45}{\includegraphics{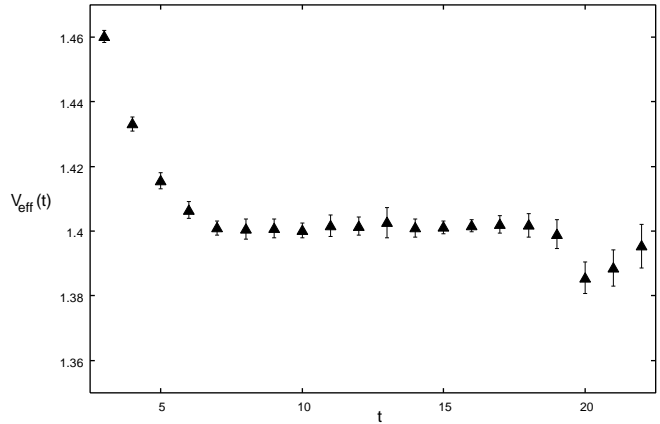}} \caption{ \label{fig8}
Effective potential as a function of $t$ in lattice units. The
graph corresponds to the internal diquark separation $D=2$  and
diquark separation $R=4$.}
\end{figure}

Finally, we compute the potential $V_{4q}$ as a function of the
distance between the two diquark, $R$, and the internal diquark
separation $D$. We calculate the ground-state potential from the
behaviour of the Wilson loop, $\langle W_{4q}\rangle$, at large
$t$ for  symmetric planar loops with $d_{1}=d_{2}=d_{3}=d_{4}=D$.
The plateau for representative values of $R$ and $D$ are shown in
Fig. \ref{fig8}. As a result of smearing the plateaus are seen at
earlier $t$, which indicates a good overlap with the state in
question. The errors on all our data points are jackknife errors.
In our analysis, we calculate the effective $4q$ potential by
choosing a fit which has $\chi^{2}/\mbox{d.o.f} \lesssim 1$ and is
insensitive to the fit parameters.

\begin{figure}[!h]
\scalebox{0.45}{\includegraphics{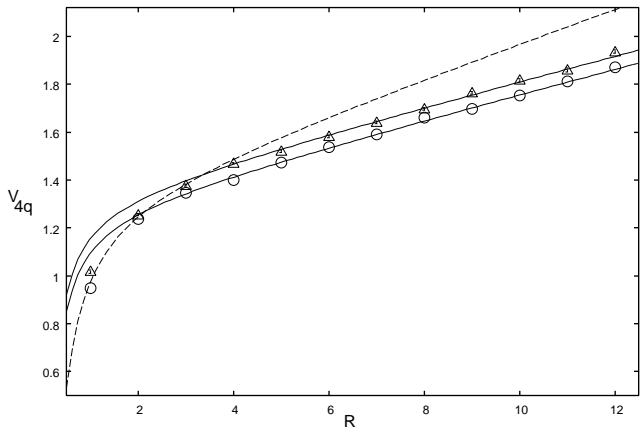}} \caption{
\label{fig9} Tetraquark potential $V_{4q}$ for symmetric planar
configurations at $D=2$ (open circles) and $D=4$ (open triangles),
respectively. The solid curve represents the one-gluon exchange
plus multi-$Y$ Ansatz, and the dashed curve the two-meson Ansatz.}
\end{figure}

Fig. (\ref{fig9}) shows the tetraquark potential as a function of
the diquark separation $R$ for different values of the internal
diquark distance $D$. At small values of diquark separation and
$R<D$, the data seem to coincide and the tetraquark potential
converges to the two-meson Ansatz. Thus for these geometries the
system corresponds to the two-meson state. For $R>D$  the
tetraquark potential is lower than $2V_{q{\bar q}}(R)$ but seems
to be in excellent agreement with one-gluon exchange Coulomb plus
multi-$Y$ Ansatz \cite{Hideo}. Comparing the $V_{4q}$ and
$2V_{q{\bar q}}$, we see that the tetraquark potential starts as a
sum of the two-meson potential and then cross  over to approach
the connected $4q$ state. This would indicate that ground-state
configuration is largely a tetraquark state for large $R$. The
flip-flop between the $4q$ state into the two-meson state  seems
nontrivial.

Now all prerequisites are available to investigate the stability
of the $4q$ state. We notice that when $qq$ and ${\bar q}{\bar q}$
are well separated, all the four quarks are linked by connected
$Y$-shaped flux tube, where the total flux-tube length is
minimised. This would mean that the examined $4q$ system can be
qualitatively explained as a connected $4q$ state. On the other
hand, when the nearest quark and antiquark are spatially close,
the system is described as a two-pion state rather than a $4q$
system. Therefore, this type of flip-flop  between the $4q$ and
the two-pion state around the cross-over can be interpreted as a
flux-tube recombination between the two mesons.

\section{Summary and conclusion}

We have presented the results of our investigation of the
tetraquark systems in improved anisotropic lattice QCD in the
quenched approximation on relatively light- and heavy-quark
masses. Our analysis takes into account all possible
uncertainties, such as statistical uncertainties, finite size, and
quenching errors when performing the chiral extrapolations. The
masses of the $J=0$ and $J=1$ states were computed using field
operators, which are motivated by the $\pi\pi$ and diquark
structure. By analysing the correlation matrix we determined the
masses of isospin $0,1$ and $2$ channels. In the region of pion
mass which we are able to access, we saw the evidence of
attraction for the $I=0$ channel. However, at relatively large
quark masses, the manifestation of the bound state changes and
$I=0$ state appears to be the scattering state.

The observation of repulsion at both light- and heavy-quark masses
is particularly evident in the $I=1$ and $I=2$ states. Both showed
a positive splitting between tetraquark and two-pion states,
suggesting repulsion as opposed to attraction. The evidence of
repulsion could be associated with the absence of a resonance in
both these channels. Moreover, in the case of $I(J^{P})=1(1^{+})$
and $0(2^{+})$, which are more exotic than the $0(0^{+})$ state,
the interpolators had sufficient overlap to allow for a successful
correlation matrix analysis and  produced no evidence of
attraction that one might expect if a tetraquark state existed.
This produces further evidence that there are no
spatially-localised lowest-lying tetraquark states.

From our static tetraquark potential analysis we conclude that we
showed that at small internal separation between diquarks and with
internal diquark distance larger than the diquark separation
($R<D$), the tetraquark system is a two-meson state and its static
potential is approximately given by  the $2V_{q\bar q}$ potential.
For $R>D$, the static potential is described by confining
$V_{4q}$. In summary, these observations suggest that, the $4q$
system behaves as a multiquark state at diquark separation larger
than the internal diquark distance. This behaviour is expected to
hold for general diquark configurations. It will be essential to
explore unquenched QCD to establish the existence or the absence
of a tetraquark resonance in lattice QCD before one rules out the
possible existence of a tetraquark state in full QCD.

\begin{acknowledgments}
We thank  R. Jaffe and C. Michael for a number of valuable
suggestions and useful comments on this work. We are grateful for
the access to the computing facility at the Shenzhen University on
128 nodes of Deepsuper-21C. ML was supported in part by the
Guangdong Provincial Ministry of Education.
\end{acknowledgments}

\end{document}